\documentclass[apj]{emulateapj}
\usepackage{apjfonts}
\submitted{Received 2009 February 25; accepted 2009 March 23; published --}
\journalinfo{Accepted to ApJL, 2009 March 23}
\usepackage{amssymb,amsmath}

\newcommand{\vcm}{V~cm$^{-1}$}
\newcommand{\ca}{Ca\,{\sc ii}~H}
\newcommand{\ha}{H$\alpha$}
\newcommand{\sm}{$\sim$}

\newcommand{\Hsi}{\textit{Reuven Ramaty High Energy Solar Spectroscopic Imager}}
\newcommand{\hsi}{\textit{RHESSI}}

\newcommand{\Soho}{\textit{Solar and Heliospheric Observatory}}
\newcommand{\soho}{\textit{SOHO}}
\newcommand{\mdi}{Michelson Doppler Imager}
\newcommand{\dg}{$^{\circ}$}
\newcommand{\hinode}{\textit{Hinode}}
\newcommand{\goes}{\textit{GOES}}
\newcommand{\GOES}{\textit{Geostationary Operational Environmental Satellite}}

\begin{document}
\title{RECONNECTION ELECTRIC FIELD AND HARDNESS OF X-RAY EMISSION OF SOLAR FLARES}
\author{Chang Liu and Haimin Wang}
\affil{Space Weather Research Lab, Center for Solar-Terrestrial Research,\\New Jersey Institute of Technology, University Heights, Newark, NJ 07102; chang.liu@njit.edu, haimin@flare.njit.edu}

\shorttitle{RECONNECTION ELECTRIC FIELD AND HARDNESS OF X-RAY EMISSION}
\shortauthors{LIU ET AL.}

\begin{abstract}
Magnetic reconnection is believed to be the prime mechanism to trigger solar flares and accelerate electrons up to energies of MeV. In the classical two-dimensional reconnection model, the separation motion of chromospheric ribbons manifests the successive reconnection that takes place higher up in the corona. Meanwhile, downward traveling energetic electrons bombard the dense chromosphere and create hard X-ray (HXR) emissions, which provide a valuable diagnostic of electron acceleration. Analyses of ribbon dynamics and HXR spectrum have been carried out separately. In this Letter, we report a study of the comparison of reconnection electric field measured from ribbon motion and hardness (spectral index) of X-ray emission derived from X-ray spectrum. Our survey of the maximum average reconnection electric field and the minimum overall spectral index for 13 two-ribbon flares show that they are strongly anti-correlated. The former is also strongly correlated with flare magnitude measured using the peak flux of soft X-ray emissions. These provide strong support for electron acceleration models based on the electric field generated at reconnecting current sheet during flares.
\end{abstract}

\section{INTRODUCTION} \label{introduction}
The ``ribbon'' structures of solar flares have long been observed at chromospheric lines (e.g, \ha\ at 656.3~nm and \ca\ at 396.8~nm). A ribbon in one magnetic polarity region has its counterpart in the other magnetic polarity region, and both run parallel to the magnetic polarity inversion line lying between them. The well-observed separation motion of ribbons is a direct mapping of energy release via magnetic reconnection in the corona reconnecting current sheet (RCS) \citep{priest02}, the rate of which can be evaluated as $\dot\phi = (\partial / \partial t) \int B\ da$, where $B$ is the corresponding magnetic field component perpendicular to the surface ribbon element $da$. Under a simplified two-dimensional reconnection model \citep{forbes84}, the electric potential drop, $V \equiv \int E\ dl$, along coronal separator with length element $dl$ equals the reconnection rate $\dot\phi$. By considering that the coronal separator has the same length as flare ribbon, the electric field can thus be expressed as $E = uB$, where u is the ribbon expansion velocity. This enables the determination of the electric field at the RCS through observable quantities in the lower atmosphere, while large uncertainties may be induced in evaluating $u$ especially for ribbons with irregular shape. Alternatively, we infer the {\it average} electric field used in the present study as
\begin{equation} \label{standard}
\langle E \rangle = \frac{\langle V \rangle}{\langle L \rangle} = \frac{\langle \dot\phi \rangle}{\langle L \rangle} = \frac{\partial}{\partial t} \left( \int B_+ \ da_+ + \Bigl\lvert \int B_- \ da_- \Bigr\rvert \right) / \left(L_+ + L_- \right) \ , \nonumber
\end{equation}
\noindent where $L$ is the length of ribbon, and the ``$+$'' and ``$-$'' subscript denote the corresponding physical quantity measured for the ribbons located in the positive and negative magnetic fields, respectively. In this way, the reconnection rate $\dot\phi$ can be calculated essentially based on progression of ribbon intensity \citep[e.g.,][]{qiu05,saba06}. Importantly, reconnecting magnetic fluxes from positive and negative magnetic fields should be identical in principle; however, their actual measurements do not always yield a good balance \citep[e.g.,][]{fletcher01}. Therefore, the average value of $E$, as defined above, may be able to represent the overall strength of reconnection electric field at the RCS, which is regarded as conceptually the most straightforward mechanism to directly accelerate electrons to high energies \citep{litvinenko96}.

A powerful diagnostic of accelerated energetic electrons produced by flares is the hard X-ray (HXR) emission, the spectrum of which often appears as a power-law distribution in photon energy [$I(\epsilon) \propto \epsilon^{-\gamma}$] \citep[e.g.,][]{tandberg88}. The hardness of X-ray emission, i.e., the power-law spectral index $\gamma$, implies a characteristic in the energy distribution of the electron flux bombarding the dense target under the bremsstrahlung emission mechanism \citep{brown71}. Specifically, a harder X-ray spectrum (with smaller spectral index) indicates that more electrons are accelerated to higher energies. For this study, X-ray spectrum is measured for entire flaring region on the solar disc, thus the derived spectral index reflects the overall hardness of X-ray emission (thereafter referred as $\langle \gamma \rangle$) created by downward traveling accelerated electrons impinging on the chromosphere.

Previous studies of temporal evolution of reconnection electric field $E$ and HXR spectral index $\gamma$ in individual flares show that they are correlated and anti-correlated, respectively, with that of the HXR flux \citep[e.g.,][]{qiu04b,grigis04}, which naturally suggests an anticorrelation relationship between $E$ and $\gamma$. This is also hinted by particle simulations with prescribed electric and magnetic fields \citep{wood05,liu+chen09}, which show that the electron energy spectrum hardens when the electric field increases. A direct comparison of $E$ and $\gamma$ using observational data is thus needed to bridge the research of magnetic reconnection and electron acceleration, which is believed to have an intrinsic causal relationship.

\begin{figure*}
\epsscale{1.}
\plotone{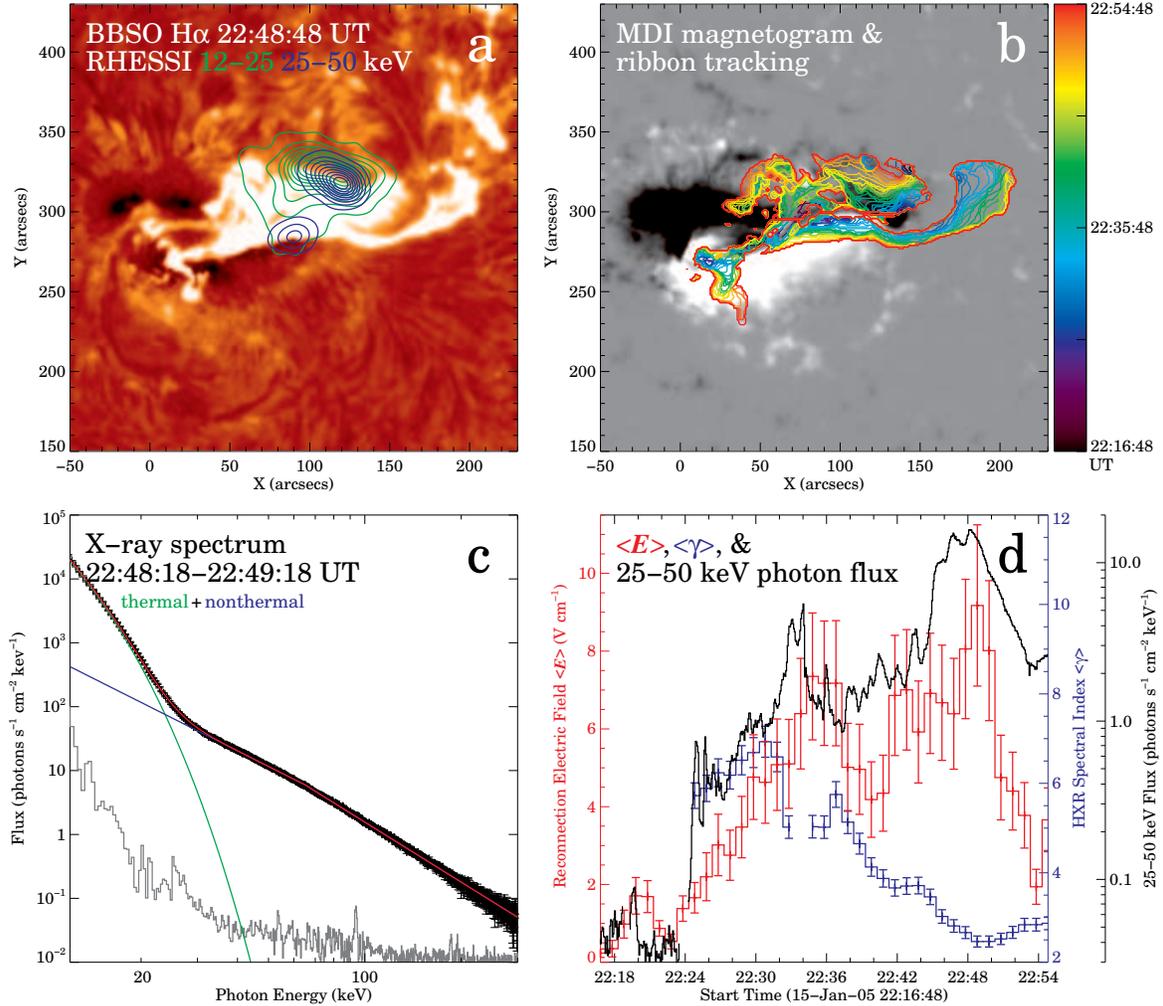}
\caption{Calculation of reconnection electric field $\langle E \rangle$ from ribbon motion and derivation of hardness of X-ray emission $\langle \gamma \rangle$ from spectral analysis for the 2005 January 15 X2.6 flare. The general flare morphology is shown in ($a$), where an \ha\ image near flare peak time is overplotted with X-ray sources. The \hsi\ images were integrated for 1 minute centered on the \ha\ image, and were reconstructed with the CLEAN algorithm using the front segments of detectors 3--9 giving a FWHM resolution of \sm9.8\arcsec. The ribbon motion tracked using a central brightness cutoff value (see \S~\ref{data}) is depicted in ($b$), overplotted on a preflare \soho\ MDI magnetogram. The X-ray spectrum ($c$) made in the same time interval as the X-ray images shows the model fitting using a thermal Maxwellian and a nonthermal double power-law components, with background flux illustrated (grey). Temporal evolution of $\langle E \rangle$, $\langle \gamma \rangle$, and \hsi\ 25--50~keV photon flux are compared in ($d$). \label{fig1}}
\end{figure*}

In retrospect, only a single such attempt has been reported. It was the 2005 May 13 M8.0 flare with ribbon-like HXR emissions \citep{liu07b}, where spatially resolved $E$ and $\gamma$ determined by tracing each position along ribbon and imaging spectroscopy for each pixel, respectively, can be compared \citep{jing07,liu08a}. As ribbon-like HXR sources are not often observed, in this Letter, we take a different approach and investigate the relationship between $\langle E \rangle$ and $\langle \gamma \rangle$ using a sample of 13 two-ribbon flares.

\section{OBSERVATIONS AND MEASUREMENTS} \label{data}
The chosen flares were all well observed on the disk with an orientation cosine factor of \sm0.7 ($\lesssim$46\dg\ from the disc center), exhibiting clear motion of ribbon separation with nearly full coverage of X-ray observations from the \Hsi\ \citep[\hsi;][]{lin02}. Ribbon observations at \ha\ and \ca\ spectral line were acquired by the Global \ha\ Network \citep{steinegger00} (10 events), the Solar Optical Telescope \citep[SOT;][]{tsuenta08} on \hinode\ (2 events), and the USAF/Optical Solar Patrol Network \citep[OSPAN, formerly known as ISOON;][]{neidig98} (1 event). These images have a cadence of 0.5--2 minutes and a pixel resolution of \sm0.1--1\arcsec. For the sake of consistency, we use photospheric magnetic fields of the flaring region (\sm2\arcsec~pixel$^{-1}$) measured with the \mdi\ \citep[MDI;][]{scherrer95} on the \Soho\ (\soho) for all the events.

\begin{deluxetable*}{lcccccccc}
\tablewidth{0pt}
\tablecaption{INFORMATION OF FLARES IN THIS STUDY\label{tbl1}}
\tablehead{
\colhead{} & \colhead{NOAA} & \colhead{Location} & \colhead{Cosine} & \colhead{\goes} & \colhead{\goes\ 1--8~\AA} & \colhead{Maximum Electric} & \colhead{Minimum HXR} & \colhead{Time Difference\tablenotemark{a}}\\
\colhead{Event Date} & \colhead{AR} & \colhead{(degree)} & \colhead{Factor} & \colhead{Level} & \colhead{Peak (UT)} & \colhead{Field $\langle E \rangle$ (\vcm)} & \colhead{Spectral Index $\langle \gamma \rangle$} & \colhead{(minutes)}
}
\startdata
(1) 2002 Feb 18 & 09830 & S20 E16 & 0.9  & M1.0 & 21:15 & 2.3 $\pm$ 0.7 & 6.84 & 0\\
(2) 2002 Feb 20 & 09830 & S18 W11 & 0.93 & M2.4 & 21:07 & 5.8 $\pm$ 1.3 & 3.64 & 0\\ 
(3) 2003 Apr 23\tablenotemark{b} & 10338 & N20 W22 & 0.87 & M2.0 & 15:56 & 3.2 $\pm$ 0.8 & 5.1 & -1\\
(4) 2003 May 27 & 10365 & S07 W17 & 0.95 & X1.3 & 23:07 & 7.7 $\pm$ 1.8 & 2.7 & 0\tablenotemark{c}\\
(5) 2003 Oct 29\tablenotemark{b} & 10486 & S15 W02 & 0.96 & X10.0 & 20:49 & 16.7 $\pm$ 5.9 & 1.88 & 1\\
(6) 2004 Mar 30 & 10581 & S05 E01 & 0.99 & C2.0 & 23:08 & 1.6 $\pm$ 0.4 & 6.94 & 0\\
(7) 2004 Nov 4\tablenotemark{b}  & 10696 & N08 E18 & 0.94 & M5.4 & 23:09 & 6.2 $\pm$ 1.3 & 2.98 & 0\\
(8) 2005 Jan 15 & 10720 & N15 W05 & 0.96 & X2.6 & 23:02 & 9.2 $\pm$ 2.1 & 2.46 & 0\\
(9) 2005 May 13 & 10759 & N12 E12 & 0.96 & M8.0 & 16:57 & 5.7 $\pm$ 1.4 & 3.21 & -1.5\\
(10) 2005 May 26\tablenotemark{b} & 10767 & S06 E13 & 0.97 & C8.6 & 21:39 & 2.1 $\pm$ 0.5 & 6.06 & 1\\
(11) 2006 Jul 6 & 10898 & S11 W32 & 0.83 & M2.5 & 08:36 & 4.9 $\pm$ 1.7 & 3.46 & -3\\
(12) 2006 Dec 14 & 10930 & S06 W46 & 0.69 & X1.5 & 22:15 & 9.2 $\pm$ 2.1 & 4.0 & -1\\
(13) 2007 Jul 10 & 10963 & S07 E45 & 0.7 & C5.2 & 17:53 & 2.1 $\pm$ 0.5 & 7.73 & 0
\enddata
\tablecomments{Flare ribbon images of the events 1--4 and 6--11, 5, and 12--13 were obtained with GHN, OSPAN, and SOT, respectively.}
\tablenotetext{a}{Time difference $=$ time of maximum average electric field $-$ time of minimum overall HXR spectral index.}
\tablenotetext{b}{We only extract results from the later phase of these events, when we consider there exhibits classical separation motion of ribbons.}
\tablenotetext{c}{\hsi\ most probably missed the main HXR peak of this event.}
\end{deluxetable*}

We take the 2005 January 15 X2.6 flare as an example to demonstrate our method of data reduction. Figure~\ref{fig1}$a$ shows an \ha\ image near the flare peak time superimposed with X-ray sources. This is a typical two-ribbon flare, with part of \ha\ ribbons occupied by HXR emitting sources (25--50~keV) and a lower energy X-ray source (12--25~keV) lying in-between. We developed a standardized procedure to derive the reconnection electric field from ribbon motion. Modules of this procedure include accurate image alignment, appropriate image destretching for ground-based observation, ribbon tracking using the intensity-based binary masks method \citep[e.g.,][]{saba06}, ribbon length estimation, and registering \ha\ images on the corresponding magnetogram by multiple feature matching. We consider uncertainties in this method that mainly stem from the following aspects: (1) The choice of the cutoff value that defines the brightened ribbons. We first determine a central cutoff value at the evolving edge of the ribbon based on the difference images relative to a fixed preflare frame. We then vary this central cutoff value within $\pm$20\% and use 11 thresholds in total to characterize the brightened ribbons at each time interval, in order to have a better evaluation of the reconnection rate; (2) Magnetic field measurement. MDI level 1.8 full-disk magnetograms are used, and a largest noise level of 30~G is taken into account. Since MDI magnetogram could be affected by high energy emissions during flares \citep[e.g.,][]{qiu03}, we use magnetograms just before the flares for calculation. We note that it is known that MDI data could suffer from Zeeman saturation, and the level 1.5 data could underestimate the flux density by a factor of \sm0.64--0.69 \citep{berger03}. However, in most of our studied events, the regions swept by flare ribbons are not obviously affected by Zeeman saturation, and the values of level 1.8 calibrated field density are \sm1.6 times those of the uncalibrated level 1.5 data\footnote{\url{http://soi.stanford.edu/magnetic/Lev1.8/}}. Therefore, we do not expect the inherent limitations in MDI measurement to significantly alter out results; and (3) Ribbon length estimation. We measure the length of the brightened ribbons as defined above. While as the shape of flare ribbons may not be regular and actual flare process may take place in a more complicated three-dimensional structure (see more discussion in \S~\ref{discussion}), we take an error of 30\% in ribbon length measurement. Moreover, caution has been taken when there is a complicated ribbon dynamics, only later stage of which exhibits classical separation motion and can be regarded as complying with the standard two-dimensional model (e.g., the 2003 October 29 X10 flare; e.g., \citealt{liu06}). Thus in such cases (indicated in Table~\ref{tbl1}), we extract results only from the clear ribbon separation phase. Figure~\ref{fig1}$b$ depicts front of the X2.6 flare ribbons as color-coded lines using a central brightness cutoff value, which shows clearly the ribbon separation motion in opposite magnetic fields. The computed $\langle E \rangle$ ({\it red}) is plotted in Figure~\ref{fig1}$d$. Its peak (\sm9.2~V~cm$^{-1}$) is co-temporal with that of the HXR emission in 25-50~keV ({\it black}), which is previously found in other events \citep{qiu04b,jing05}.

For analysis of X-ray spectrum, we used the Object Spectral Executive (OSPEX) software package and applied the pileup correction, if necessary, in SolarSoftWare (SSW). We set the energy bin width to 1/3 keV, the standard finest binning, and choose the length of fitting time interval to be the same as the cadence of ribbon images centering on the time of each frame, except in the case that we have to avoid the time when \hsi\ attenuators change status. We model the thermal plasma and high-energy emissions with an isothermal component and nonthermal power-law distributions, respectively. Quite often, the nonthermal emission is best fitted using a double power-law function with two different spectral indices. In such a case, we take the spectral index reflecting the hardness of photons with medium energies (\sm20--50~keV) as the result for that event (see discussion in \S~\ref{discussion}). The uncertainty in the values of HXR spectral index may arise from the choice of the length of fitting interval and spectrum background, and is estimated to be \sm5\%. The X-ray spectral fitting results of the X2.6 flare around its peak time is shown in Figure~\ref{fig1}$c$, where $\langle \gamma \rangle$ reaches \sm2.46 in \sm20--60~keV. The time profile of $\langle \gamma \rangle$ ({\it blue}) is plotted in Figure~\ref{fig1}$d$, in comparison with those of $\langle E \rangle$ and the 25--50~keV photon flux. As usual, the minimum overall spectral index occurs when the HXR peaks, which is believed to be an intrinsic feature of electron acceleration in the flare impulsive phase \citep[e.g.,][]{grigis04}.

\section{RESULTS} \label{result}
Following similar procedure, we calculate reconnection electric field and HXR spectral index for other events, and give the results in Table~\ref{tbl1}. The maximum $\langle E \rangle$ and the minimum $\langle \gamma \rangle$ lie in the range of \sm1--20~\vcm\ and \sm2--8, respectively, which is generally comparable with what previously reported \citep[e.g.,][]{jing05,lin01}. In a majority of events, the times when $\langle E \rangle$ and $\langle \gamma \rangle$ reach the maximum and minimum values, respectively, agree within one minute.

\begin{figure}
\epsscale{1.15}
\plotone{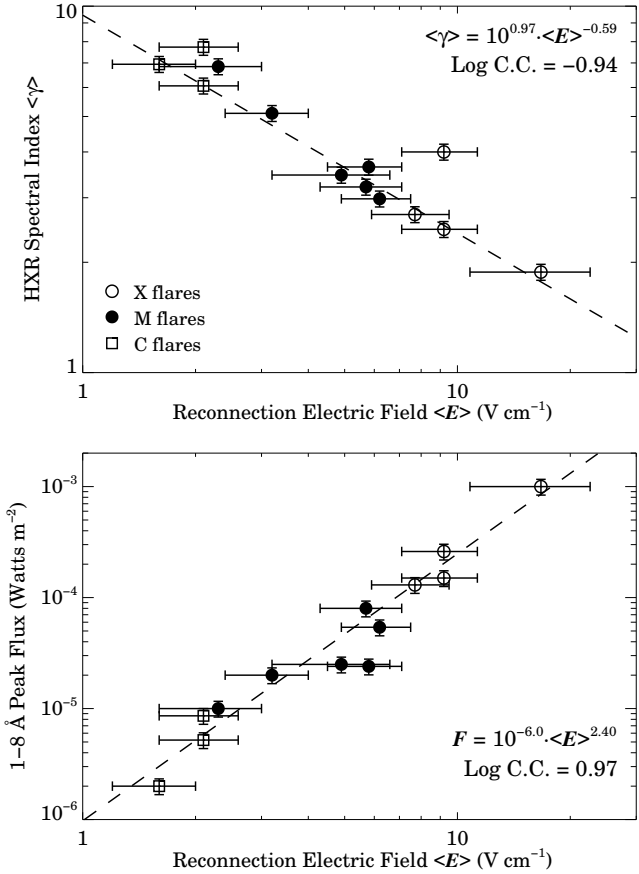}
\caption{Scatter plots of the minimum overall HXR spectral index ({\it top}) and the peak flux in 1--8~\AA\ ({\it bottom}) vs. the maximum average electric field in the RCS in a logarithmic scale. The dashed line shows the least-squares linear fit to the data points. The rms measurement error for the \goes\ 1--8~\AA\ channel is 16.2\% \citep{garcia94}. \label{fig2}}
\end{figure}

Figure~\ref{fig2} ({\it top}) shows the scatter plot of these two quantities, with a value of logarithmic correlation coefficient (CC) computed to be $-$0.94. That is to say, a larger electric field in the RCS corresponds to a harder X-ray spectrum (with smaller $\gamma$) and presumably to a harder electron precipitation spectrum, which is qualitatively consistent with the trend predicted by numerical simulations \citep{wood05,liu+chen09}. This high correlation relationship thus strongly support the hypothesis that direct acceleration by the electric field generated by magnetic reconnection may play an important role in producing energetic electrons in flares. We note that X, M, and C-class flares appear to be three populations, which indicates that flares with larger magnitude are almost always associated with larger electric fields and harder HXR spectra.

This tendency is more clearly seen in Figure~\ref{fig2} ({\it bottom}), where we make scatter plot of the maximum $\langle E \rangle$ versus the peak flux in 1--8~\AA\ soft X-ray measured by the \GOES\ \citep[\goes;][]{kahler91}. The logarithmic CC also has a high coefficient of 0.97 for all the data sets. This further implies that larger electric field strength may be responsible for stronger electron acceleration and hence stronger high-energy emissions seen in flares with larger magnitude.

\section{SUMMARY AND DISCUSSION} \label{discussion}
Joining the research of magnetic reconnection and electron acceleration, we have determined the average electric field $\langle E \rangle$ in the RCS and the overall hardness of X-ray emission $\langle \gamma \rangle$ from flare observations of 13 events. We then find a strong correlation relationship between both the minimum $\langle \gamma \rangle$ and the flare magnitude corresponding to the maximum $\langle E \rangle$. This is, to our knowledge, the clearest direct observational evidence obtained so far of electron acceleration by reconnection electric field in solar flares, which has long been explored but mostly by theoretical models \citep{aschwanden05}. We discuss the present results with related observation and simulation works.

The present research is closest to that of \citet{liu08a}, who compared spatially resolved electric field and HXR spectral index for a single event. In their result, an anticorrelation relationship between $E$ and $\gamma$ is also found, while for the $E$ with similar strength, the corresponding $\gamma$ generally has smaller values than that in our study. A possible explanation for this discrepancy is that different from the average $E$ and the overall $\gamma$ herein presented, they explored the local electric field by tracing motion of every position along ribbon, and local spectral index in each pixel by applying an imaging spectroscopy technique. A larger error could be induced in calculating such $E$ and $\gamma$, but it is more possible that different approach, i.e., global or local, matters in studying the flare energy release. In another study, \citet{ning08} found that $\dot \phi$ is anti-correlated with $\gamma$ in two flares. However, one of their event is close to limb (with a cosine factor of \sm0.49), and they used ribbon observations in 1600~\AA\ that forms in upper chromosphere and transition region. More importantly, instead of $\dot \phi$, $E$ is believed to be the physical quantity directly associated with electron acceleration.

Previous studies of reconnection electric field usually use the form, $E = uB$, while $u$ is not easy to be measured accurately \citep[cf.][]{qu04}. We use another form, $E = \dot \phi / L$, where $\dot \phi$ can be evaluated in detail using ribbon intensity \citep[e.g.,][]{saba06}. Here we must note that the ribbon length $L$ is an invariable in the standard two-dimensional model; however, it usually involves rapidly during actual flares especially in the early phase. Although chromospheric ribbons may be regarded as an instantaneous response of energy release in the coronal current sheet through precipitation of accelerated electrons, recent observational studies have suggested the requirement of consideration of three-dimensional magnetic structures \citep[e.g.,][]{liu07b,jing07}. The deviation from the standard two-dimensional model is also implied by the asymmetry of flare geometry. For the studied events, the ratio of reconnection electric field and ribbon length measured in opposite magnetic field regions, i.e., $(\phi_+ L_+^{-1})/(\phi_- L_-^{-1})$ and $L_+/L_-$, generally falls into the range of \sm0.2--2 and 0.3--3, respectively. Without an applicable three-dimensional flare model, we thus use the equation in \S~\ref{introduction} as an approximation, where we measure the length of ribbon when it fully developed and treat it as a constant. This simplification may be justified, because most of our events do not show a significant change of ribbon length at times when the maximum $\langle E \rangle$ and the minimum $\langle \gamma \rangle$ occur. Using the time-dependent change of ribbon length as a probe to the magnetic reconnection process is an area of on-going research \citep[e.g.,][]{lee08}.

Our observational results demonstrate that there exists a remarkable anticorrelation relationship between $E$ and $\gamma$. This qualitatively agrees with model results \citep{wood05,liu+chen09}, in that both indicate a hardening of electron energy distribution with increasing electric field strength. Quantitatively comparison is beyond the scope of this study, as further assumptions need to be made before numerical results of electron spectra can be compared with observed photon spectra. An important issue is that in the simulation of \citet{liu+chen09}, it is found that the accelerated electrons exhibit not only a power-law energy spectrum but also an exponential tail at the high energy end. The latter indicates that diffuse shock acceleration may be at work. Therefore we take only the spectral indices of the HXRs with medium energies as the results, when the spectrum can not be fitted using a single power-law function. This also implies that different acceleration mechanisms may all contribute to electron acceleration during flares \citep{aschwanden05}. Nevertheless, the presence of such clear correlations of observational signatures provides strong support for electron acceleration models based on electric field generated at the RCS.

\acknowledgments
The authors thank the teams of GHN, \hinode, OSPAN, \hsi, and \soho\ for efforts in obtaining the data. We thank the referee for helpful comments that improved the Letter. C.~L. is indebted to Na Deng for valuable help in programming and discussion. C.~L. and H.~W. were supported by NSF grants ATM 08-19662 and ATM 07-45744, and NASA grants NNX 08AQ90G and NNX 07AH78G. GHN is operated by the Space Weather Research Lab, New Jersey Institute of Technology. \hinode\ is a Japanese mission developed and launched by ISAS/JAXA, with NAOJ as domestic partner and NASA and STFC (UK) as international partner. It is operated by these agencies in co-operation with ESA and NSC (Norway). OSPAN is a PI driven project by Air Force Research Laboratory Space Vehicles Directorate (RVBXS) and the National Solar Observatory. \hsi\ is a NASA Small Explorer. \soho\ is a project of international cooperation between ESA and NASA.


\begin{thebibliography}{31}
\expandafter\ifx\csname natexlab\endcsname\relax\def\natexlab#1{#1}\fi

\bibitem[{{Aschwanden}(2005)}]{aschwanden05}
{Aschwanden}, M.~J. 2005, Physics of the Solar Corona (New York: Springer)

\bibitem[{{Berger} \& {Lites}(2003)}]{berger03}
{Berger}, T.~E., \& {Lites}, B.~W. 2003, \solphys, 213, 213

\bibitem[{{Brown}(1971)}]{brown71}
{Brown}, J.~C. 1971, \solphys, 18, 489

\bibitem[{{Fletcher} \& {Hudson}(2001)}]{fletcher01}
{Fletcher}, L., \& {Hudson}, H. 2001, \solphys, 204, 69

\bibitem[{{Forbes} \& {Priest}(1984)}]{forbes84}
{Forbes}, T.~G., \& {Priest}, E.~R. 1984, in Solar Terrestrial Physics: Present
  and Future, ed. D.~M. {Butler} \& K.~{Papadopoulous}, Vol.~1, 35--39

\bibitem[{{Garcia}(1994)}]{garcia94}
{Garcia}, H.~A. 1994, \solphys, 154, 275

\bibitem[{{Grigis} \& {Benz}(2004)}]{grigis04}
{Grigis}, P.~C., \& {Benz}, A.~O. 2004, \aap, 426, 1093

\bibitem[{{Jing} {et~al.}(2007){Jing}, {Lee}, {Liu}, {Gary}, \&
  {Wang}}]{jing07}
{Jing}, J., {Lee}, J., {Liu}, C., {Gary}, D.~E., \& {Wang}, H. 2007, \apjl,
  664, L127

\bibitem[{{Jing} {et~al.}(2005){Jing}, {Qiu}, {Lin}, {Qu}, {Xu}, \&
  {Wang}}]{jing05}
{Jing}, J., {Qiu}, J., {Lin}, J., {Qu}, M., {Xu}, Y., \& {Wang}, H. 2005, \apj,
  620, 1085

\bibitem[{{Kahler} \& {Kreplin}(1991)}]{kahler91}
{Kahler}, S.~W., \& {Kreplin}, R.~W. 1991, \solphys, 133, 371

\bibitem[{{Lee} \& {Gary}(2008)}]{lee08}
{Lee}, J., \& {Gary}, D.~E. 2008, \apjl, 685, L87

\bibitem[{{Lin} {et~al.}(2001){Lin}, {Feffer}, \& {Schwartz}}]{lin01}
{Lin}, R.~P., {Feffer}, P.~T., \& {Schwartz}, R.~A. 2001, \apjl, 557, L125

\bibitem[{{Lin} {et~al.}(2002)}]{lin02}
{Lin}, R.~P., {et~al.} 2002, \solphys, 210, 3

\bibitem[{{Litvinenko}(1996)}]{litvinenko96}
{Litvinenko}, Y.~E. 1996, \apj, 462, 997

\bibitem[{{Liu} {et~al.}(2006){Liu}, {Lee}, {Deng}, {Gary}, \& {Wang}}]{liu06}
{Liu}, C., {Lee}, J., {Deng}, N., {Gary}, D.~E., \& {Wang}, H. 2006, \apj, 642,
  1205

\bibitem[{{Liu} {et~al.}(2008){Liu}, {Lee}, {Jing}, {Gary}, \& {Wang}}]{liu08a}
{Liu}, C., {Lee}, J., {Jing}, J., {Gary}, D.~E., \& {Wang}, H. 2008, \apjl,
  672, L69

\bibitem[{{Liu} {et~al.}(2007){Liu}, {Lee}, {Yurchyshyn}, {Deng}, {Cho},
  {Karlick{\'y}}, \& {Wang}}]{liu07b}
{Liu}, C., {Lee}, J., {Yurchyshyn}, V., {Deng}, N., {Cho}, K.-S.,
  {Karlick{\'y}}, M., \& {Wang}, H. 2007, \apj, 669, 1372

\bibitem[{{Liu} {et~al.}(2009){Liu}, {Chen}, {Ding}, \& {Fang}}]{liu+chen09}
{Liu}, W.~J., {Chen}, P.~F., {Ding}, M.~D., \& {Fang}, C. 2009, \apj, 690, 1633

\bibitem[{{Neidig} {et~al.}(1998)}]{neidig98}
{Neidig}, D., {et~al.} 1998, in ASP Conf. Ser., Vol. 140, Synoptic Solar
  Physics, ed. K.~S. {Balasubramaniam}, J.~{Harvey}, \& D.~{Rabin}, 519--528

\bibitem[{{Ning}(2008)}]{ning08}
{Ning}, Z. 2008, \apss, 314, 137

\bibitem[{{Priest} \& {Forbes}(2002)}]{priest02}
{Priest}, E.~R., \& {Forbes}, T.~G. 2002, \aapr, 10, 313

\bibitem[{{Qiu} \& {Gary}(2003)}]{qiu03}
{Qiu}, J., \& {Gary}, D.~E. 2003, \apj, 599, 615

\bibitem[{{Qiu} {et~al.}(2004){Qiu}, {Wang}, {Cheng}, \& {Gary}}]{qiu04b}
{Qiu}, J., {Wang}, H., {Cheng}, C.~Z., \& {Gary}, D.~E. 2004, \apj, 604, 900

\bibitem[{{Qiu} \& {Yurchyshyn}(2005)}]{qiu05}
{Qiu}, J., \& {Yurchyshyn}, V.~B. 2005, \apjl, 634, L121

\bibitem[{{Qu} {et~al.}(2004){Qu}, {Shih}, {Jing}, \& {Wang}}]{qu04}
{Qu}, M., {Shih}, F., {Jing}, J., \& {Wang}, H. 2004, \solphys, 222, 137

\bibitem[{{Saba} {et~al.}(2006){Saba}, {Gaeng}, \& {Tarbell}}]{saba06}
{Saba}, J.~L.~R., {Gaeng}, T., \& {Tarbell}, T.~D. 2006, \apj, 641, 1197

\bibitem[{{Scherrer} {et~al.}(1995){Scherrer}, {Bogart}, {Bush}, {Hoeksema},
  {Kosovichev}, {Schou}, {Rosenberg}, {Springer}, {Tarbell}, {Title},
  {Wolfson}, {Zayer}, \& {MDI Engineering Team}}]{scherrer95}
{Scherrer}, P.~H., {Bogart}, R.~S., {Bush}, R.~I., {Hoeksema}, J.~T.,
  {Kosovichev}, A.~G., {Schou}, J., {Rosenberg}, W., {Springer}, L., {Tarbell},
  T.~D., {Title}, A., {Wolfson}, C.~J., {Zayer}, I., \& {MDI Engineering Team}.
  1995, \solphys, 162, 129

\bibitem[{{Steinegger} {et~al.}(2000){Steinegger}, {Denker}, {Goode},
  {Marquette}, {Varsik}, {Wang}, {Otruba}, {Freislich}, {Hanslmeier}, {Luo},
  {Chen}, \& {Zhang}}]{steinegger00}
{Steinegger}, M., {Denker}, C., {Goode}, P.~R., {Marquette}, W.~H., {Varsik},
  J., {Wang}, H., {Otruba}, W., {Freislich}, H., {Hanslmeier}, A., {Luo}, G.,
  {Chen}, D., \& {Zhang}, Q. 2000, in ESA Spec. Publ., Vol. 463, The Solar
  Cycle and Terrestrial Climate, Solar and Space weather, ed. A.~{Wilson},
  617--622

\bibitem[{{Tandberg-Hanssen} \& {Emslie}(1988)}]{tandberg88}
{Tandberg-Hanssen}, E., \& {Emslie}, A.~G. 1988, {The Physics of Solar Flares}
  (Cambridge and New York: Cambridge Univ. Press)

\bibitem[{{Tsuneta} {et~al.}(2008){Tsuneta}, {Ichimoto}, {Katsukawa}, {Nagata},
  {Otsubo}, {Shimizu}, {Suematsu}, {Nakagiri}, {Noguchi}, {Tarbell}, {Title},
  {Shine}, {Rosenberg}, {Hoffmann}, {Jurcevich}, {Kushner}, {Levay}, {Lites},
  {Elmore}, {Matsushita}, {Kawaguchi}, {Saito}, {Mikami}, {Hill}, \&
  {Owens}}]{tsuenta08}
{Tsuneta}, S., {Ichimoto}, K., {Katsukawa}, Y., {Nagata}, S., {Otsubo}, M.,
  {Shimizu}, T., {Suematsu}, Y., {Nakagiri}, M., {Noguchi}, M., {Tarbell}, T.,
  {Title}, A., {Shine}, R., {Rosenberg}, W., {Hoffmann}, C., {Jurcevich}, B.,
  {Kushner}, G., {Levay}, M., {Lites}, B., {Elmore}, D., {Matsushita}, T.,
  {Kawaguchi}, N., {Saito}, H., {Mikami}, I., {Hill}, L.~D., \& {Owens}, J.~K.
  2008, \solphys, 249, 167

\bibitem[{{Wood} \& {Neukirch}(2005)}]{wood05}
{Wood}, P., \& {Neukirch}, T. 2005, \solphys, 226, 73

\end{thebibliography}
\end{document}